\documentstyle[art12]{article}
\begin{document}
\begin{center}
{\large \bf HEAVY QUARK FRAGMENTATION FUNCTIONS
 IN THE HEAVY QUARK  EFFECTIVE THEORY}
\end{center}
\begin{center}
{\bf Martynenko A.P., Saleev V.A.\\
\it Samara State University, Samara 443011, Russia}
\end{center}
\begin{abstract}

We calculate fragmentation functions for a b-quark to fragment into
color-singlet P-wave bound states $\bar c b$ in the Heavy Quark Effective
Theory with the exact account of $O(1/m_b)$ corrections. We demonstrate
an agreement of the obtained results with the corresponding calculations
carried out in quantum chromodynamics.
\end{abstract}

\vspace{3mm}
{\Large \bf Introduction.}

\vspace{2mm}
The study of heavy quarkonia properties is a subject of much current interest
for understanding of quark-gluon interaction dynamics. $B_c$-mesons, consisting
of b- and c-quarks, hold a unique position in the heavy quarkonia physics
\cite{1}.  In the first place, $B_c$ -mesons consist of two heavy quarks, so
the predictions of potential models refer to $J/\Psi$- and $\Upsilon$ -mesons
as well as to $B_c$ -mesons \cite{2}. In the second place, $B_c$ -mesons are
constructed from the quarks of different flavors and masses, what essentially
determine their decay characteristics \cite{3}.

Production of mesons with heavy quarks in $e^+e^-$, $\gamma\gamma$ and
$p\bar  p$-interactions may be described in nonrelativistic perturbative
quantum chromodynamics. At present, two mechanisms were investigated for the
production of $B_c$-mesons: the recombination and the fragmentation.
In the first
case, $B_c$-mesons are formed from heavy quarks, produced independently in
hard subprocess. The fragmentational mechanism demands the pair production of
$b$- or $c$-quarks in hard subprocess with their subsequent fragmentation to
$B_c$-mesons ($\bar b\to B_c \bar c, c\to B_c b$). The relative contributions
of these mechanisms in the production cross-sections are different in various
reactions. In $e^+e^-$ -annihilation only quark fragmentation is essential
\cite{4}. In $p\bar p$ É $\gamma\gamma$ -interactions the fragmentational
mechanism prevails also for $B_c$-mesons production with large transverse
momenta \cite{5,6}. In the range of small transverse momenta the recombination
is dominant and determines the total cross section of $B_c$-meson production
in $\gamma\gamma$ and $p\bar p$ interactions. But experimental conditions of
$B_c$-meson discovery are more perspective in the large transverse momenta
domain. So, the study of heavy quark fragmentation into $B_c$-mesons
attracts considerable interest. An approach for the calculation of
fragmentation
functions $D_{\bar b\to  B_c  \bar  c},  D_{c\to B_c b}$ in nonrelativistic
perturbative quantum chromodynamics was suggested in \cite{7}.

At the same time, in the last years there was suggested, based on QCD, the
Heavy
Quark Effective Theory (HQET) \cite{8,9} for description of
heavy hadrons properties. HQET make it possible to obtain the finite
analytical result with some accuracy even for complicated processes of quark-
gluon interaction. In this approach the matrix elements of different processes
may be decomposed on degrees of two small parameters: the strong coupling
constant $\alpha_s(m_Q)$ and $\Lambda_{QCD}/m_Q$, where $m_Q$ is the mass of
the heavy quark. In the limit $m_Q\rightarrow\infty$ the effective lagrangian,
which describes the strong interactions of heavy quarks has an exact
spin-flavor
symmetry \cite{8}. HQET is successfully used for investigation of exclusive and
inclusive hadron decays \cite{9}.
Recently it was shown \cite{10} that HQET may be used for study of
b-quark fragmentation into S-wave pseudoscalar and vector mesons and the
corresponding nonpolarized fragmentation functions were calculated.
The HQET calculation of the b-quark fragmentation into the transverse and
longitudinal polarized S-wave $B_c^*$-mesons have been made in \cite{10b}.
In this
work we have calculated the fragmentation functions of b-quark into P-wave
color-singlet states ($\bar c b$) to the next to leading order in the heavy
quark mass expansion using the methods of HQET.

\section{Fragmentation functions into P-wave $\bar c b$-mesons.}

 Heavy b-quark may fragment into bound states of two heavy quarks ($\bar c b$ -
states) with orbital momentum l=1. There are four such states: $^1P_1,
^3P_J$ (J=0,1,2). Heavy quark fragmentation functions into P-wave $B_c$-
mesons
were calculated by Chen \cite{11} and Yuan \cite{12} in QCD, but the results
of their calculations disagree. Let carry out the similar calculation of
fragmentation functions in the HQET. Let $q=m_b v+k$ is 4-momentum of virtual
heavy quark, $p_1=(1-r)Mv+\rho$ and $p_2=rMv-\rho$ are 4-momenta of $b-$ and
$\bar c$-quarks correspondingly; $\rho$ is 4-momentum of relative motion. Let
also $l=k-\rho$ is 4-momentum of the virtual gluon and k is the residual
momentum of the fragmenting heavy quark. Fragmentation functions for the
process
$b\rightarrow B_c+c$ are determined by the next expression \cite{7}:
\begin{equation}
D(z)=\frac{1}{16\pi^2}\int ds \theta\left(s-\frac{M^2}{z}-\frac{m_c^2}
{1-z}\right)\lim_{q_0\rightarrow\infty}\frac{\sum\vert T\vert^2}{\sum\vert
T_0\vert^2},
\end{equation}
where $M=m_b+m_c$ is the mass of $B_c$-meson, T is the matrix element
for production $B_c+\bar c$ from an off-shell $b^\ast$-quark with virtuality
$s=q^2$, and $T_0$ is the matrix element for producing an on-shell b-quark
with the same 3-momentum $\vec q$. The calculation can be greatly simplified
by using the axial gauge with gauge parameter $n_\mu=(1,0,0,-1)$ in the frame
where $q_\mu=(q_0,0,0,\sqrt{q_0^2-s})$:

\begin{equation}
D_{\sigma\lambda}(k)=\frac{1}{k^2+i0}\left[g_{\sigma\lambda}-\frac{k_\sigma
n_\lambda+
k_\lambda n_\sigma}{k\cdot n}+\frac{n^2 k_\sigma k_\lambda}{(k\cdot
n)^2}\right],
\end{equation}

The part of amplitude T that involves production of the virtual $b^\ast$-quark
can be treated as an unknown Dirac spinor $\Gamma$. In the limit
$q_0\rightarrow
\infty$, the same spinor factor $\Gamma$ appears in the matrix element
$T_0=\bar\Gamma v(q)$, what leads to cancellation of this factor $\Gamma$ in
(1).

Let consider the fragmentation of b-quark into color-singlet bound state
($\bar c b$) $^1P_1$. The amplitude of such process involves the spinor factor
$v(p_1)\bar u(p_2)$. To project the pair of quarks on $^1P_1$ bound state we
have used the next substitution \cite{13}:
\begin{equation}
v(p_2)\bar u(p_1)\rightarrow\sqrt{M}\frac{\hat p_2-m_c}{2m_c}\gamma_5\frac{\hat
p_1+m_b}{2m_b}.
\end{equation}

The HQET Lagrangian, including the leading and the $1/m_b$ terms is given by
\cite{8,9}:
\begin{equation}
L=\bar h_v\left\{iv\cdot D+\frac{1}{2m_b}\left[C_1(iD)^2-C_2(v\cdot iD)^2-
\frac{C_3}{2}g_s\sigma_{\mu\nu}G^{\mu\nu}\right]\right\}h_v,
\end{equation}
where
\begin{equation}
C_1=1,~~C_2=3\left(\frac{\alpha_s(\mu)}{\alpha_s(m_b)}\right)^{-\frac{8}
{(33-2n_f)}}-2,~~C_3=\left(\frac{\alpha_s(\mu)}{\alpha_s(m_b)}\right)^{-\frac{9}
{(33-2n_f)}}.
\end{equation}
All of these coefficients are equal to 1 at the heavy quark mass scale
$\mu=m_b$.
When we concerned the fragmentation into P-wave mesons, it is necessary to
decompose the projecting operator (3) and the gluon propagator (2) on the
relative motion momentum $\rho$. Using the Feynman rules, derived from HQET
Lagrangian (4), we may write the full fragmentation amplitude into $^1P_1$
-state \cite{13}:
\begin{equation}
iM(n^1P_1)=\frac{\sqrt{4\pi
M}\alpha_s}{3m_cm_b}R'_1(0)\epsilon_\alpha(L_z)\frac{\partial }
{\partial \rho_\alpha}\Big\{\frac{1}{l^2}(-g_{\mu\nu}+\frac{n_\nu l_\mu}{n\cdot
l})
\end{equation}
\begin{displaymath}
\Bigl\{\bar u(p')\gamma^\nu(m_c \hat v-\hat\rho-m_c)\gamma_5(m_b\hat
v+\hat\rho+m_b)
[v^\mu+\frac{C_1}{2m_b}(\rho+k)^\mu-
\end{displaymath}
\begin{displaymath}
-\frac{C_2}{2m_b}v\cdot(\rho+k)v^\mu+i\frac{C_3}{2m_b}\sigma^{\mu\lambda}
(\rho-k)_\lambda]\frac{1+\hat v}{2}\Gamma\frac{i}{v\cdot k+\frac{C_1}{2m_b}k^2-
\frac{C_2}{2m_b}(v\cdot k)^2}\Bigr\}\vert_{\rho=0},
\end{displaymath}
where $\epsilon_\alpha(L_z)$ is the polarization vector of $^1P_1$ -state.
To calculate amplitude (6) it is convenient to divide it into two parts on the
degrees of small parameter $1/m_b$. When $m_b\rightarrow\infty$ in vertex
function and in the propagator of heavy quark, we obtain the main contribution
to
the fragmentation amplitude of b-quark in the form:
\begin{equation}
iM_1(n^1P_1)=\frac{\sqrt{4\pi M}\alpha_s
2R'_1(0)}{3r^2(s-m_b^2)^3}\epsilon^\ast_\alpha
(L_z)\bar u(p')W^\alpha\gamma_5\Gamma,
\end{equation}
\begin{displaymath}
W_\alpha=(s-m_b^2)\left[(\hat v+1)\gamma_\alpha-\frac{v\cdot k}{n\cdot k}\hat n
(\hat v+1)\gamma_\alpha\right]+4mk_\alpha\left[1+\frac{k\cdot v}{k\cdot n}\hat
n
\right](\hat v-1)-
\end{displaymath}
\begin{displaymath}
-2Mr(s-m_b^2)v_\alpha\frac{1}{n\cdot k}\hat n(\hat v-1)+2Mr(s-m_b^2)n_\alpha
\frac{v\cdot k}{(n\cdot k)^2}\hat n(\hat v-1).
\end{displaymath}

All calculations of the fragmentation functions, which are rather complicated,
were done by means of the system "REDUCE". Substituting (7) into (1), we
obtain:
\begin{equation}
D_1(n^1P_1)(y)=N_1\frac{(1-y)^2}{ry^8}(9y^4-4y^3+40y^2+96),~~~N_1=\frac{\alpha_s^2
\vert R'_{nP}(0)\vert^2}{54\pi r^5M^5}.
\end{equation}
where $y=(1-z+rz)/rz$ is the so called Yaffe-Randall variable \cite{14}, and
$r=m_c/M$.

The amplitude of bound $\bar c b$ state $n^3P_J$ production may be derived from
(6), changing $\gamma_5\rightarrow \hat\epsilon(S_z)$, where
$\epsilon^\mu(S_z)$ is the spin wave function:
\begin{equation}
iM(n^3P_J)=\frac{\sqrt{4\pi
M}\alpha_s}{3m_cm_b}R'_1(0)\sum_{S_z,L_z}<1,L_z;1,S_z\vert J,J_z>
\epsilon^\ast_\beta(S_z)\epsilon^\ast_\alpha(L_z)
\end{equation}
\begin{displaymath}
\frac{\partial }
{\partial \rho_\alpha}\Bigl\{\frac{1}{l^2}(-g_{\mu\nu}+\frac{n_\nu
l_\mu}{n\cdot l})
\bar u(p')\gamma^\nu(m_c \hat v-\hat\rho-m_c)\gamma_\beta(m_b\hat
v+\hat\rho+m_b)
[v^\mu+\frac{C_1}{2m_b}(\rho+k)^\mu-
\end{displaymath}
\begin{displaymath}
-\frac{C_2}{2m_b}v\cdot(\rho+k)v^\mu+i\frac{C_3}{2m_b}\sigma^{\mu\lambda}
(\rho-k)_\lambda]\frac{1+\hat v}{2}\Gamma\frac{i}{v\cdot k+\frac{C_1}{2m_b}k^2-
\frac{C_2}{2m_b}(v\cdot k)^2}\Bigr\}\vert_{\rho=0},
\end{displaymath}
where we have expressed the Clebsch-Gordon coefficients and $\epsilon^\ast_
\beta(S_z), \epsilon^\ast_\alpha(L_z)$ by the bound state polarizations
\cite{13}:
\begin{equation}
\sum_{S_z,L_z}<1,L_z;1,S_z\vert J,J_z>\epsilon^\ast_\beta(S_z)
\epsilon^\ast_\alpha(L_z)=\Bigl\{
\begin{array}{l}
\frac{1}{\sqrt{3}}(g_{\alpha\beta}-v_\alpha v_\beta),~~~J=0\\
\frac{i}{\sqrt{2}}\epsilon_{\alpha\beta\lambda\rho}v_\lambda\epsilon^\ast_\rho(J_z),~~~J=1\\
\epsilon_{\alpha\beta}(J_z),~~~~~~~~~~~~~~J=2.
\end{array}
\end{equation}

Taking in (9) only the terms of leading order on $1/m_b$ and doing necessary
differentiation on $\rho_\alpha$, we have obtained:
\begin{equation}
iM_1(n^3P_J)=\frac{\sqrt{4\pi
M}\alpha_s2R'_1(0)}{3r^2(s-m_b^2)^3}\sum_{S_z,L_z}<1,L_z;1,S_z\vert J,J_z>
\epsilon^\ast_\beta(S_z)\epsilon^\ast_\alpha(L_z)
\end{equation}
\begin{displaymath}
\bar u(p')\Bigl\{(s-m_b^2)(1-\frac{k\cdot v}{k\cdot n}\hat
n)\gamma_\alpha\gamma_\beta-
4k_\alpha M(1+\frac{k\cdot v}{k\cdot n}\hat n)\gamma_\beta
-4rM^2\left(\frac{k\cdot v}{k\cdot n}\right)^2n_\alpha\hat n\gamma_\beta\Bigr\}
(1+\hat v)\Gamma.
\end{displaymath}
This amplitude determines the basic contribution to fragmentation functions
of b-quark into $^3P_J$- state:
\begin{eqnarray}
D_1(n^3P_0)(y)=N_1\frac{(y-1)^2}{ry^8}(y^4-4y^3+8y^2+32),\\
D_1(n^3P_1)(y)=N_1\frac{2(y-1)^2}{ry^8}(3y^4-4y^3+16y^2+48),\\
D_1(n^3P_2)(y)=N_1\frac{20(y-1)^2}{ry^8}(y^4+4y^2+8).
\end{eqnarray}

The heavy quark fragmentation functions into P-wave mesons were calculated
in \cite{12} using the QCD. Our results (8), (12)-(14) coincide with the
calculations of \cite{12} in the leading order on $1/m_b$. Let consider
calculation of the next-to-leading order contributions $O(1/m_b)$ for
$^1P_1$ -state. First of all, let observe, that the fragmentation functions,
determined by the amplitudes (9) and (11), involves factor $1/[1+r(y-1)]$.
Decompose it on degrees of r, we have the correction to (8):
\begin{equation}
D_2(n^1P_1)(y)=N_1\frac{(1-y)^3}{y^8}(9y^4-4y^3+40y^2+96),
\end{equation}

Without any calculations we can obtain the propagator correction $O(1/m_b)$
to (8), which arises as a result of following substitution:
\begin{equation}
\frac{1}{v\cdot k}\rightarrow\frac{1}{v\cdot
k+\frac{C_1}{2m_b}k^2-\frac{C_2}{2m_b}
(v\cdot k)^2}\approx\frac{1}{v\cdot k}\left[1+r(-C_1+\frac{1}{2}C_2My)\right].
\end{equation}
This correction has the next form:
\begin{equation}
D_3(n^1P_1)(y)=N_1\frac{(y-1)^2}{y^8}(9y^4-4y^3+40y^2+96)(-2C_1+C_2My).
\end{equation}
The vertex correction of necessary order in (6) is calculated in a more
tedious way:
\begin{equation}
iM_{vert.}(n^1P_1)=\frac{\sqrt{4\pi
M}\alpha_s}{3m_cm_b}R'_{nP}(0)\epsilon^\ast_
\alpha(L_z)\frac{\partial}{\partial\rho_\alpha}\Bigl\{\bar u(p')\gamma_\nu(m_c
\hat v-\hat\rho-m_c)\gamma_5
\end{equation}
\begin{displaymath}
(m_b\hat v+\hat\rho+m_b)
\left[\frac{C_1}{2m_b}(\rho+k)_\mu-\frac{C_2}{2m_b}v\cdot(\rho+k)v_\mu+i
\frac{C_3}{2m_b}\sigma_{\mu\lambda}(k-\rho)_\lambda\right]
\end{displaymath}
\begin{displaymath}
\frac{1+\hat v}{2}\Gamma\frac{i}{v\cdot k}\frac{1}{l^2}
\left(-g_{\mu\nu}+\frac{n_\nu l_\mu}{n\cdot l}
\right)\Bigr\}\vert_{\rho=0}.
\end{displaymath}
It is natural to perform it as a sum of several terms. Putting $\rho=0$ in
the square brackets, we obtain:
\begin{equation}
\left[\frac{C_1}{2m_b}(\rho+k)_\mu-\frac{C_2}{2m_b}v\cdot(\rho+k)v_\mu-
\frac{C_3}{4m_b}[\gamma_\mu(\hat\rho-\hat k)-(\hat\rho-\hat
k)\gamma_\mu]\right]\vert_{\rho=0}=
\end{equation}
\begin{displaymath}
=\left[\frac{C_1}{2m_b}k_\mu-\frac{C_2}{2m_b}(v\cdot k)v_\mu+\frac{C_3}
{4m_b}(\gamma_\mu\hat k-\hat k\gamma_\mu)\right].
\end{displaymath}
An addendum $(-C_2(k\cdot v)v_\mu/2m_b)$ in (19) gives the contribution to
fragmentation functions which differs only by sign from the similar quark
propagator correction:
\begin{equation}
D_4(n^1P_1)(y)=N_1\frac{(y-1)^2}{y^8}(9y^4-4y^3+40y^2+96)(-C_2My).
\end{equation}
Two other terms in square brackets of (19) give rise the next fragmentation
amplitudes:
\begin{equation}
iM^{(1)}_{vert.}(n^1P_1)=\frac{\sqrt{4\pi M}2\alpha_s}{3m_cm_b}R'_{nP}(0)
\epsilon^\ast_\alpha(L_z)\frac{C_1}{2m_b}\frac
{1}{r(s-m_b^2)^2}
\bar u(p')\Bigl[2k_\alpha Mr\hat n\frac{(s-m_b^2)}{k\cdot n}-
\end{equation}
\begin{displaymath}
 -4k_\alpha m\hat k+
2n_\alpha\hat n Mr^2\frac{(s-m_b)^2}{(k\cdot n)^2}
+r\hat n\gamma_\alpha\frac{(s-m_b^2)^2}{k\cdot n}-\hat k\gamma_\alpha(s-m_b^2)
\Bigr]\gamma_5(1+\hat v)\Gamma,
\end{displaymath}
\begin{equation}
iM^{(2)}_{vert.}(n^1P_1)=\frac{\sqrt{4\pi M}2\alpha_s}{3m_cm_b}R'_{nP}(0)
\epsilon^\ast_\alpha(L_z)\frac{C_3}{m_b}\frac{1}{r^2(s-m_b^2)^2}
\end{equation}
\begin{displaymath}
\bar u(p')\Bigl\{-2k_\alpha Mr\hat n\frac{(s-m_b^2)}{k\cdot n}-8k_\alpha M\hat
k-
2(s-m_b^2)k_\alpha+2Mr\frac{(s-m_b^2)}{k\cdot n}\hat n\gamma_\alpha\hat k+
\end{displaymath}
\begin{displaymath}
+r\hat n\gamma_\alpha\frac{(s-m_b^2)^2}{k\cdot n}+r(s-m_b^2)\hat k\gamma_\alpha
\Bigr\}\gamma_5(1+\hat v)\Gamma.
\end{displaymath}
A further part of vertex correction $O(1/m_b)$ appears, when we differentiate
the expression in square brackets of (18) on $\rho_\alpha$:
\begin{equation}
iM^{(3)}_{vert.}(n^1P_1)=\frac{\sqrt{4\pi M}\alpha_s}{3m_cm_b}R'_{nP}(0)
\epsilon^\ast_\alpha(L_z)\frac{1}{2r(s-m_b^2)^2}
\end{equation}
\begin{displaymath}
\bar u(p')\Bigl[C_1\left(k_\alpha\hat n\frac{1}{k\cdot n}-\gamma_\alpha\right)
+C_3\Bigl(k_\alpha\hat n\frac{1}{k\cdot n}+\frac{1}{2M}\frac{(s-m_b^2)}{k\cdot
n}
\hat n\gamma_\alpha-
\end{displaymath}
\begin{displaymath}
\frac{1}{k\cdot n}\hat n\hat k\gamma_\alpha+2\gamma_\alpha\Bigr)
\Bigr]\gamma_5(1+\hat v)\Gamma.
\end{displaymath}

The contributions of matrix elements (21)-(23) to fragmentation functions have
the following form:
\begin{equation}
D_5(n^1P_1)(y)=4N_1\frac{(y-1)^2y}{y^8}(3y^3-2y^2+6y+32)C_1,
\end{equation}
\begin{equation}
D_6(n^1P_1)(y)=4N_1\frac{(y-1)y^2}{y^8}(-3y^3-4y^2+14y-16)C_3,
\end{equation}
\begin{equation}
D_7(n^1P_1)(y)=2N_1\frac{(y-1)^2y}{y^8}\left[(3y^3+6y^2+4y+32)C_1+2y
(3y^2-2y-16)C_3\right].
\end{equation}

The small component of heavy quark field also leads to the correction of type
O(r) in function D(y). Substituting the small quark component propagator to (6)
instead of heavy quark propagator
\begin{equation}
\frac{i}{v\cdot k}\frac{1+\hat v}{2}\left[\frac{1}{2m_b}\sigma_{\mu\lambda}
k^\lambda\right]\frac{1-\hat v}{2},
\end{equation}
we obtain the next amplitude of $B_c$- meson production:
\begin{equation}
iM_{prop.}(n^1P_1)=-\frac{\sqrt{4\pi M}2\alpha_s}{3m_cm_b}R'_{nP}(0)
\epsilon^\ast_\alpha(L_z)\frac{4}{r^2(s-m_b^2)^3}
\end{equation}
\begin{displaymath}
\bar u(p')\left[4k_\alpha M\hat k+2k_\alpha(s-m_b^2)+r\frac{(s-m_b^2)^2}
{k\cdot n}\hat n\gamma_\alpha+(s-m_b^2)\hat k\gamma_\alpha\right]
\gamma_5(1-\hat v)\Gamma.
\end{displaymath}
This amplitude gives the contribution to fragmentation function of the kind:
\begin{equation}
D_8(n^1P_1)(y)=6N_1\frac{(y-1)y}{y^8}(y^5+2y^4+3y^3+16y-16).
\end{equation}

The calculation of b-quark fragmentation functions into $^3P_J$ states was done
in a similar way. The results of $D(n^3P_J)$ calculations are represented in
the next section.

\section{Discussion of the results.}

 Let analyse b-quark fragmentation functions into P-wave mesons. As was
mentioned earlier, in the leading order on $1/m_b$ the expressions (8), (12)-
(14) coincide with QCD calculations of Yuan \cite{12}. The correction $O(r)$
to (8) is determined by the sum of terms (15), (17), (20), (24)-(26), (29).
Setting $C_1=C_3=1$, we obtain the fragmentation function $D(n^1P_1)$ to the
next-to-leading order on r:
\begin{equation}
D(n^1P_1)(y)=\frac{\alpha_s^2\vert R'_{nP}(0)\vert^2}{54\pi r^5M^5}
\frac{(y-1)^2}{ry^8}\Bigl[(9y^4-4y^3+40y^2+96)-
\end{equation}
\begin{displaymath}
-r(3y^5-31y^4+32y^3+8y^2-192y+96)\Bigr].
\end{displaymath}
Similarly, for the $^3P_J$ states, we have:
\begin{equation}
D(^3P_0)(y)=\frac{\alpha_s^2\vert R'_{nP}(0)\vert^2}{54\pi r^5M^5}
\frac{(y-1)^2}{ry^8}\Bigl[(y^4-4y^3+8y^2+32)+
\end{equation}
\begin{displaymath}
+\frac{r}{3}(3y^5-11y^4+392y^2+192y-96)\Bigr],
\end{displaymath}
\begin{equation}
D(^3P_1)(y)=\frac{\alpha_s^2\vert R'_{nP}(0)\vert^2}{27\pi r^5M^5}
\frac{(y-1)^2}{ry^8}\Bigl[(3y^4-4y^3+16y^2+48)+
\end{equation}
\begin{displaymath}
+r(3y^5+5y^4+8y^3+32y^2+96y-48)\Bigr],
\end{displaymath}
\begin{equation}
D(^3P_2)(y)=\frac{10\alpha_s^2\vert R'_{nP}(0)\vert^2}{27\pi r^5M^5}
\frac{(y-1)^2}{ry^8}\Bigl[(y^4+4y^2+8)+
\end{equation}
\begin{displaymath}
+\frac{r}{15}(-3y^5-y^4+36y^3-164y^2+240y-120)\Bigr].
\end{displaymath}
It follows immediately from (30)-(33), that our results coincide with the
calculations of Yuan \cite{12} with the accuracy $O(r)$, if we take into
account, that the nonperturbative factor of (30)-(33) in \cite{12} contains
reduced mass $\mu$ contrary to our factor $rM$. Using obtained fragmentation
functions (30)-(33), we may calculate the fragmentation probabilities of
corresponding $(\bar c b)$-mesons \cite{7}:
\begin{equation}
P_{b\rightarrow \bar c b}(n^1P_1)=\int_0^1 dz D_{b\rightarrow \bar c b}
(n^1P_1)(z,r)=
\end{equation}
\begin{displaymath}
=\frac{\alpha_s^2\vert R'_{nP}(0)\vert^2}{54\pi r^5M^5}\Bigl[
\frac{r\ln r}{(1-r)^8}\left(21+32r+110r^2-184r^3-387r^4-168r^5\right)+
\end{displaymath}
\begin{displaymath}
+\frac{1}{210(1-r)^7}\left(1032+4497r+21353r^2-2762r^3-65202r^4-76199r^5-
3679r^6\right)\Bigr].
\end{displaymath}
\begin{equation}
P_{b\rightarrow \bar c b}(n^3P_0)
=\frac{\alpha_s^2\vert R'_{nP}(0)\vert^2}{54\pi r^5M^5}\Bigl[
\frac{r\ln r}{3(1-r)^8}\left( 3+8r+98r^2+1496r^3-1221r^4-960r^5\right)+
\end{equation}
\begin{displaymath}
+\frac{1}{630(1-r)^7}\left(360+1767r-14977r^2+180778r^3+158658r^4-427697r^5-
19849r^6\right)\Bigr].
\end{displaymath}
\begin{equation}
P_{b\rightarrow \bar c b}(n^3P_1)
=\frac{\alpha_s^2\vert R'_{nP}(0)\vert^2}{27\pi r^5M^5}\Bigl[
\frac{r\ln r}{(1-r)^8}\left(3+16r+10r^2+112r^3-237r^4-192r^5\right)+
\end{equation}
\begin{displaymath}
+\frac{1}{210(1-r)^7}\left(348-177r+8419r^2+2714r^3+14334r^4-81769r^5-
4349r^6\right)\Bigr].
\end{displaymath}
\begin{equation}
P_{b\rightarrow \bar c b}(n^3P_2)
=\frac{10\alpha_s^2\vert R'_{nP}(0)\vert^2}{27\pi r^5M^5}\Bigl[
\frac{r\ln r}{15(1-r)^8}\left(33-68r+130r^2-548r^3-291r^4+24r^5\right)+
\end{equation}
\begin{displaymath}
+\frac{1}{3150(1-r)^7}\left(1710+2697r-2525r^2-10330r^3-146760r^4+3425r^5+
583r^6\right)\Bigr].
\end{displaymath}
Putting here $\vert R'_{2P}(0)\vert^2=0.201$ GeV$^5$, $m_c=1.5$ GeV,
 $m_b=4.9$ GeV
and $\alpha_s(2m_c)$=0.38 ($2m_c$ is a minimal energy of exchanged gluon), we
have obtained the numerical value of the fragmentation probabilities, which
are presented in table. We see, that our integral probabilities of P-wave
$\bar c b$ meson production, founded by means of the b-quark fragmentation
functions in the HQET are in good agreement with the results of QCD
calculations
of \cite{12}.\\[3mm]

\begin{tabular}{|c|c|c|c|}   \hline
$P_{b\rightarrow \bar cb}(2^1P_1)$&$P_{b\rightarrow \bar cb}(2^3P_0)$&
$P_{b\rightarrow\bar cb}(2^3P_1)$&
$P_{b\rightarrow\bar c b}(2^3P_2)$   \\   \hline
$6.4\cdot 10^{-5}$  &$2.5\cdot 10^{-5}$ &$7.3\cdot 10^{-5}$ &$10.5\cdot
10^{-5}$       \\  \hline
\end{tabular}
\vspace{5mm}

So the performed calculations show that the Heavy Quark Effective Theory may
be successfully used for the study of the heavy quark fragmentation. In this
approach we may systematically take into account the $O(1/m_b)$ corrections
in the amplitudes and the probabilities of the fragmentation, what increases
the accuracy of HQET calculations. Moreover, it seems more important, that
HQET leads to finite analytical answer, when we study complicated problems in
the heavy quark physics. The approach, based on HQET, may be used for
calculation of heavy quark fragmentation functions into D-wave mesons, and
for the investigation of $B_c$ -meson hadroproduction.

We are grateful to Faustov R.N., Kiselev V.V., Likhoded A.K. for useful
discussions of $B_c$ meson physics and the Heavy Quark Effective Theory, and
to Braaten E., Cheung K., Fleming S. for the valuable information about
obtained results.

This work was done under the financial support of the Russian Fund of
Fundamental
Researches (Grant 93-02-3545) and by State Committee on High Education of
Russian Federation (Grant 94-6.7-2015).


\begin{thebibliography}{99}
\bibitem{1}Gershtein S.S., et al. //UFN. 1995. V.165. N1. P.3.
\bibitem{2}Gershtein S.S., et al.//Yad. Fiz. 1988. V.48. P.515.
\bibitem{3}Kiselev V.V., Tkabladze A.V. //Yad. Fiz. 1988. V.48. P.515.
\bibitem{4}Kiselev V.V., Likhoded A.K., Shevlyagin M.V. //Yad.Fiz. 1994.
V.57. N4. P.733\\
Chang C.-H, Chen Y.-Q //Physics Letters. 1992. B284. P.127.
\bibitem{5}Berezhnoy A.V., Likhoded A.K., Shevlyagin M.V. //Preprint IHEP
94-48 1994; 94-82 1994; 95-59 1995.
\bibitem{6}Kolodziej K., Leike A., Ruckl R. //Preprint MPI-PhT/94-84;
MPI-PhT/95-36; Phys.Lett. 1995. V.B348. P.219.
\bibitem{7}Braaten E., Cheung K., Yuan T.C. //Phys. Rev. 1993. V.D48. N11.
P.R5049\\
Cheung K., Yuan T.C. //Physics Letters. 1994. V.B325. P.481\\
Braaten E., Yuan T.C. //Phys. Rev. Letters. 1993. V.71. N11. P.1673\\
\bibitem{8}Neubert M. //Physics Reports. 1994. V.245. P.261.
\bibitem{9}Grinstein B. //Preprint UCSD/PTH 94-24.
\bibitem{10}Braaten E., et al. //Preprint FERMILAB-PUB-94-305-T 1994.
\bibitem{10b} Martynenko A.P., Saleev V.A. // Izvest.Vuzov.Fizika (in Russian),
in publ.
\bibitem{11}Chen Y.-Q.//Physical Review. 1993. V.D48. P.5158.
\bibitem{12} Yuan T.C. //Preprint UCD-94-2 1994.
\bibitem{13}Kuhn J.H., Kaplan J., Safiani El G.O. //Nuclear Physics. 1979.
V.B157. P.125.
\bibitem{14}Jaffe R., Randall L. //Nuclear Physics. 1994. V.B412. P.79
\end{thebibliography}
\end{document}